\documentclass[sigconf]{acmart}
\AtBeginDocument{%
  }

\copyrightyear{2026}
\acmYear{2026}
\setcopyright{cc}
\setcctype{by}
\acmConference[LLM4Code '26]{3rd International Workshop on Large Language Models For Code }{April 12--18, 2026}{Rio de Janeiro, Brazil}
\acmBooktitle{3rd International Workshop on Large Language Models For Code (LLM4Code '26), April 12--18, 2026, Rio de Janeiro, Brazil}
\acmPrice{}
\acmDOI{10.1145/3786181.3788724}
\acmISBN{979-8-4007-2412-1/2026/04}

\usepackage{amsmath}
\usepackage{graphicx}
\usepackage{textcomp}
\usepackage{xcolor}
\usepackage{todonotes}
\usepackage{subfig}
\usepackage{float}
\usepackage{url}
\usepackage{algorithm}
\usepackage{algpseudocode}

\begin{document}

\title{Code Roulette: How Prompt Variability Affects LLM Code Generation}


\author{Andrei Paleyes}
\affiliation{%
  \institution{University of Cambridge}
  \city{Cambridge}
  \state{England}
  \country{UK}
}
\email{ap2169@cl.cam.ac.uk}

\author{Radzim Sendyka}
\affiliation{%
  \institution{University of Cambridge}
  \city{Cambridge}
  \state{England}
  \country{UK}
}
\email{rs2071@cam.ac.uk}

\author{Diana Robinson}
\affiliation{%
  \institution{University of Cambridge}
  \city{Cambridge}
  \state{England}
  \country{UK}
}
\email{dmpr3@cl.cam.ac.uk}

\author{Christian Cabrera}
\affiliation{%
  \institution{University of Cambridge}
  \city{Cambridge}
  \state{England}
  \country{UK}
}
\email{chc79@cam.ac.uk}

\author{Neil D. Lawrence}
\affiliation{%
  \institution{University of Cambridge}
  \city{Cambridge}
  \state{England}
  \country{UK}
}
\email{ndl21@cam.ac.uk}


\begin{abstract}
Code generation is one of the most active areas of application of Large Language Models (LLMs). While LLMs lower barriers to writing code and accelerate development process, the overall quality of generated programs depends on the quality of given prompts. Specifically, functionality and quality of generated code can be sensitive to user's background and familiarity with software development. It is therefore important to quantify LLM's sensitivity to variations in the input. To this end we propose an evaluation pipeline for LLM code generation with a focus on measuring sensitivity to prompt augmentations, completely agnostic to a specific programming tasks and LLMs, and thus widely applicable. We provide extensive experimental evidence illustrating utility of our method and share our code for the benefit of the community.
\end{abstract}



\maketitle

\section{Introduction}
Since the emergence of Large Language Models (LLMs), one of their most widespread applications is code generation \cite{zhang2023unifying}. Ability of LLMs to generate code has the potential to bring many advantages, such as lowering the barriers to coding, democratising and accelerating software development \cite{robinson2024requirements}. Notably, requirements for generated code are formulated by end users in natural language, often lacking any particular structure. In a traditional software development process software engineers collect such requirements from different stakeholders and generate a system specification that satisfies end-user requirements. It is a trust-based relationship between users and engineers.

With the onset of LLMs people with all kinds of backgrounds and coding expertise can use a chat interface to generate code directly. However, at the present moment of LLM development we cannot trust these models as we would trust another human being, because we cannot be certain the model understands our requirements. An important step to build such trust is to make LLMs safer and better understand their behaviour in the context of code generation. In this paper we describe a procedure to aid such understanding, specifically to measure sensitivity of a LLM code generation to variability in the input prompt.

Investigating how sensitive LLM code outputs are to differences in input prompts is important because it will determine what kind of actions are needed to produce the same code outputs from varying prompt inputs that describe the same requirement. For example, we may need to provide more guidance to users, such as follow-up questions as they provide the prompts, as part of the coding pipeline. Or this might involve techniques such as generating other related inputs or other related outputs based on distance and averaging them together to regularise the output code generated. Testing how sensitive LLM outputs are to changes in inputs allows us to understand the scope of this challenge to standardise code outputs to varying descriptions of the same requirement. This paper makes strides towards addressing this challenge. Specifically, we put forth following contributions:
\begin{enumerate}
    \item An evaluation procedure to measure LLM's sensitivity on code generation tasks.
    \item A sensitivity analysis evaluation of multiple popular LLMs.
    \item Openly available code that other researches can use to replicate our work and build on it\footnote{\url{https://github.com/apaleyes/code-gen-sensitivity}}.
\end{enumerate}

\section{Motivation}
Intuitively, our work is motivated by the observation that people can express their thoughts differently depending on their background, education, experience and other contexts. This notion includes the way humans write prompts to interact with LLMs. In this section we present evidence from the literature confirming this observation.

Recent research by Chen et al. \cite{Chen2024-nd} investigated robustness of code LLMs to variations that could be expected in real world application based on a literature review and survey of professionals who use LLMs for code. They identified 18 categories of these perturbations such as extra space outside of words. They highlight an example of a prompt for LLM code generation tool, StarCoder, where a tiny change to the prompt with the repetition of the word ``it'' resulted in incorrect code generated. We anticipate much larger prompt variability from differences in how people decompose problems that they are trying to build code for based on their background knowledge and experience, including what technical concepts about programming they have available to them in their mental models of both programming languages and LLMs.

There is significant evidence that decomposing a problem and thinking about it in a way that can be translated into code is a nontrivial skill. As discussed below, work in computational thinking and research into teaching programming highlights this.

Computational thinking is a type of analytical thinking that draws on concepts fundamental to computing \cite{Wing2008-re}. It involves choosing the right abstraction to describe a problem in a way that can be solved by a computer. This style of thinking is much discussed in educational contexts as a way of helping people to build or understand these solutions and apply them in other context (see for example \cite{Neumann2021-ju}).

Problem decomposition is an important skill in programming, as a way of formulating subgoals to implement in code \cite{Poitras2024-wa}. As LLMs become intertwined in the process of generating code, educators are adapting their courses to incorporate them. For example, Vadaparty et al. designed a CS1 course for programming using LLMs which changed the emphasis from syntax toward explaining and testing code generated from LLMs and decomposing complex problems into small pieces that can be used to craft prompts for an LLM \cite{Vadaparty2024-pi}. Of their many learning goals, they identified as essential problem decomposition, even if instructors only adopt part of the course.

Programmers rely on mental models of programs and programming concepts to guide their work \cite{Heinonen2023-ac}. As programming shifts to LLMs, mental models will also shift. Some work that illustrates this is by Liang et al. \cite{Liang2024-eh}, who found that prompt programming differed from traditional software development in that programmers were relying on mental models of LLMs rather than mental models of code. Ehsani et al. explored where developers understanding of LLM responses can come apart and developed a tool to offer templates to developers to improve prompts and enable LLM-driven issue resolution \cite{Ehsani2025-cy}.

We can also see this effect of differences in background and knowledge impacting prompt development in multidisciplinary software teams. For example, in a study on prototyping with generative AI tools that discussed differences in approaches to creating prompts between team members such as UX researchers, AI engineers, and product managers \cite{Subramonyam2024-ow}.

While differences in problem decomposition and mental models between users with varying backgrounds represent an important area for future investigation, this work focuses on a more fundamental question: how sensitive are LLMs to basic variations in prompt text that preserve the same underlying requirement? As a first step toward understanding prompt variability, we examine sensitivity to simple textual perturbations (typos, synonyms, and paraphrasing) that any user might introduce regardless of their background. In the following sections we confirm the sensitivity of LLM generated code to synthetic changes to prompts, with important implications for development of code with LLMs.

\section{Methodology}
We designed an evaluation pipeline to investigate the sensitivity of LLM generated code to fluctuations in the prompts. In this section we describe this pipeline and motivate key design decisions behind it.

\subsection{Overview of the pipeline}
Let $\mathcal{P}$ denote the set of all possible prompts representing coding tasks, and let $\mathcal{M}: \mathcal{P} \rightarrow \mathcal{C}$ be a function that maps a prompt to generated code, where $\mathcal{C}$ represents the space of all code snippets. In the context of this work, $\mathcal{M}$ is a LLM.

To systematically evaluate sensitivity, we define an augmentation function $\mathcal{F}: \mathcal{P} \times [0,1] \rightarrow \mathcal{P}$ that perturbs a rate $r \in [0,1]$ of the input prompt, thereby producing a modified prompt. Additionally, we employ a distance function $\mathcal{D}: \mathcal{C} \times \mathcal{C} \rightarrow [0,1]$ that quantifies the dissimilarity between two code snippets, yielding a normalized distance metric. Specific realizations of both $\mathcal{F}$ and $\mathcal{D}$ are discussed in subsequent sections. While our current experiments use lexical augmentations (typos, synonyms, paraphrasing), the pipeline is designed to accommodate more sophisticated augmentation methods that model differences in problem decomposition or conceptual framing.

The evaluation procedure operates as follows. Given a single prompt $p \in \mathcal{P}$, we first establish a reference baseline by generating $n$ independent code samples from the unperturbed prompt. These reference outputs serve as ground truth representations of the model's behavior on the original task specification. We then systematically vary the augmentation rate parameter from zero to one in discrete increments. The variation in the augmentation rate $r$ generates code outputs that are progressively more perturbed versions of the original prompt. For each augmentation level, we compute the pairwise distances between all generated outputs and the reference baseline set, subsequently aggregating these measurements to obtain an overall distance metric. This metric quantifies how the model's output distribution changes as a function of prompt augmentation rate, thus providing insight into the model's sensitivity to input variations. Formal description of the pipeline is given in Algorithm~\ref{algo:pipeline}.

\begin{algorithm}
\caption{Evaluation Pipeline for Sensitivity of LLM Code Generation}
\label{algo:pipeline}
\begin{algorithmic}[1]
\Require Prompt $p$, model $\mathcal{M}$, augmentation function $\mathcal{F}$, distance function $\mathcal{D}$, number of reference samples $n$, augmentation rate step $\delta$
\State Initialize empty set $R \leftarrow \emptyset$
\State Generate reference code set: $C_{ref} \leftarrow \{c_i = \mathcal{M}(p) : i = 1, \ldots, n\}$
\For{$r \leftarrow 0$ to $1$ step $\delta$}
    \State Generate augmented code set: $C_{aug} \leftarrow \{c_j = \mathcal{M}(\mathcal{F}(p, r)) : j = 1, \ldots, n\}$
    \State Initialize distance sum: $S \leftarrow 0$
    \State Initialize count: $N \leftarrow 0$
    \For{each $c_{ref} \in C_{ref}$}
        \For{each $c_{aug} \in C_{aug}$}
            \State Compute pairwise distance: $d \leftarrow \mathcal{D}(c_{ref}, c_{aug})$
            \State $S \leftarrow S + d$
            \State $N \leftarrow N + 1$
        \EndFor
    \EndFor
    \State Compute average distance: $\bar{d}_r \leftarrow S / N$
    \State Store result: $R \leftarrow R \cup \{(r, \bar{d}_r)\}$
\EndFor
\State \Return $R$
\end{algorithmic}
\end{algorithm}

\subsection{Data}\label{subsection:data}
We use three datasets for evaluation experiments.

\textbf{LeetCode (Old)} dataset consists of programming tasks from the popular website LeetCode. LeetCode tasks is a popular benchmark for code generation, and is widely used in scientific literature. We have compiled a set of 20 tasks randomly sampled from this benchmark\footnote{The particular instance of the dataset we used is published on HuggingFace at \url{https://huggingface.co/datasets/NyanDoggo/leetcode}}. Unfortunately the most commonly circulated LeetCode tasks are included in the training data of most modern LLMs, which means they achieve almost 100\% correct performance on these problems. This issue is known in the literature as ``data contamination'', and as shown by Sendyka et al. \cite{sendyka2025llmperformancecodegeneration} does not improve even with complete obfuscation of text.

\textbf{LeetCode (New)} dataset is the list of 20 programming tasks posted on LeetCode in March 2025, that we have collected manually. Matching their publication dates and official training cut-off dates of the LLMs we used, we have high confidence these tasks were not used in the training data for LLM models we evaluated.

\textbf{Our Dataset}\label{label:data} is the collection of 22 programming tasks we have created ourselves, spanning topics of simulations, algorithms, data science, application development, and games. We made a deliberate effort to specify details that would separate our tasks from common programming problems found online. Problems in this dataset do not have strictly one correct answer, and are designed to be open ended, which is significantly different from LeetCode exercises. The dataset is fully available in the repository accompanying this paper\footnote{Full dataset is available at \url{https://github.com/apaleyes/code-gen-sensitivity/blob/main/Sensitivity\%20of\%20LLMs\%20tasks\%20dataset.md}}.

\subsection{Prompt Augmentation methods}
We apply three augmentation methods to the prompts: keyboard typos, synonyms, and paraphrasing.

\textbf{Keyboard typos} augmentation randomly replaces $rate$ characters in the prompt with an adjacent key on the QWERTY keyboard. This method attempts to simulate typos that a human can make while typing text on a keyboard. We use the NLPaug library to implement this method \cite{ma2019nlpaug}.

\textbf{Synonyms} augmentation randomly replaces $rate$ words in the prompt with their synonyms. Synonyms are defined by semantic meaning using the WordNet database \cite{miller1995wordnet}. This method is also implemented using NLPaug.

\begin{figure}
    \centering
    \includegraphics[width=0.45\textwidth]{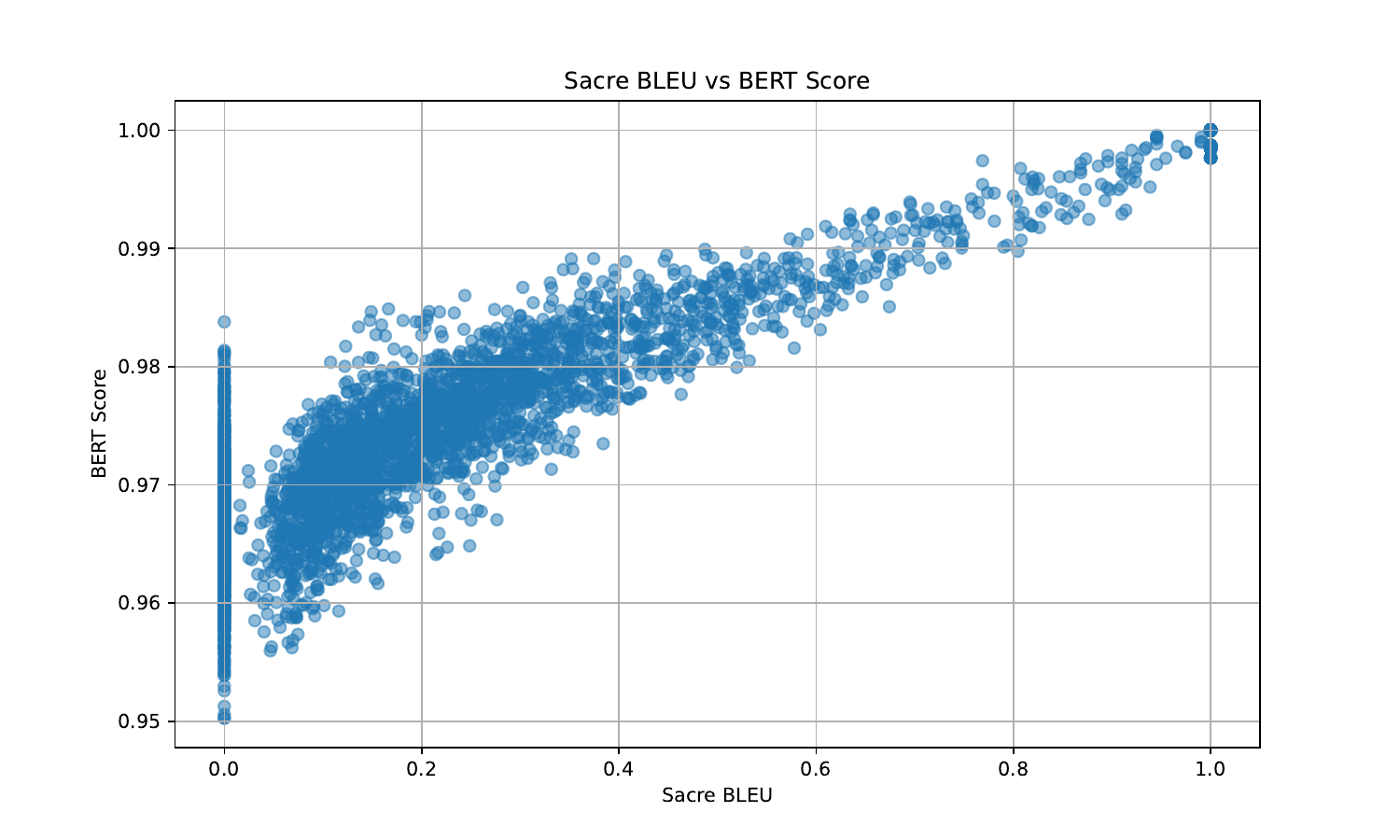}
    \caption{Diversity (i.e., Sacre BLEU) versus semantic similarity (i.e., BERT Score) of the generated paraphrases against the original prompt for Our Dataset. The generated paraphrases are semantically similar, while textually diverse.}
    \label{fig:paraphrasing}
\end{figure}

\textbf{Paraphrasing} uses the translation capabilities of LLMs (i.e., Gemini) to paraphrase the prompt. The process aims to generate phrases that maintain the semantic similarity of the prompt while varying their vocabulary, thereby achieving text diversity. The resulting paraphrases are used to evaluate the sensitivity of the LLMs when generating code for diverse prompts that are semantically similar. Figure~\ref{fig:paraphrasing} plots the BERT Score~\cite{zhang2020bertscore} (i.e., semantic similarity) versus the Sacre BLEU~\cite{post-2018-call} (i.e., text diversity) metric for the generated paraphrases for the prompts in Our Dataset. The generated paraphrases are semantically similar and diverse, with BERT score varying between 0.95 and 1.0 and Sacre BLEU metric varying between 0 and 1.0. A lower Sacre BLEU metric value indicates higher diversity.

\subsection{Output distance measures}
We use the Tree Similarity of Edit Distance (TSED) metric to compute the distance between two LLM code outputs. TSED is a recently published improvement over syntax tree edit distance with an openly available implementation \cite{song-etal-2024-revisiting}. TSED works as a function $t(c_1, c_2) \rightarrow d \in [0, 1]$, where $c_1$ and $c_2$ are fragments of code and $d$ is a real value between $0$ and $1$ indicating similarity between $c_1$ and $c_2$, $0$ meaning completely different and $1$ meaning identical. Importantly, we do not evaluate correctness of generated code, and specifically focus on deviations between the codes generated from the original and altered prompts. The motivation behind this decision is that even if the code produced by a model is incorrect, it still represents a possible output the user might see, and therefore is a valid data point when measuring LLM sensitivity.

By its definition TSED measures syntactic differences, not semantic equivalence or functional correctness. Structurally different code may produce identical outputs, while structurally similar code may behave differently. In this work our focus is on measuring LLM output consistency rather than correctness. Even on its own structural variability matters: users providing similar prompts may receive dramatically different implementations, affecting code review, maintenance, and trust, even when the code is functionally equivalent. Future work should complement structural analysis with functional testing to distinguish benign diversity from problematic instability.

\begin{figure}
    \centering
    \subfloat{\includegraphics[width=0.45\textwidth]{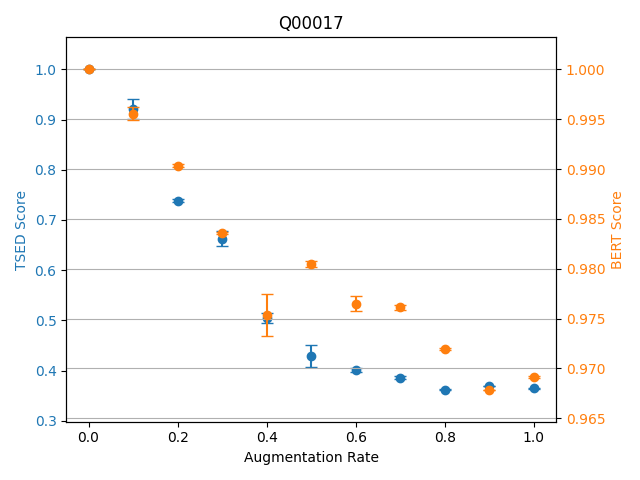}}
    \caption{Comparison of BERT and TSED metrics on an example task. While overall trend is comparable, TSED uses most of its possible range, while BERT only varies between 0.96 and 1.0.}
    \label{fig:tsed_vs_bert}
\end{figure}

It is also worth noting that we are not using BLEU (quantifying n-grams overlap), BERT score (cosine similarity in BERT contextual embeddings space) or their derivatives to evaluate code similarity. We found these general-purpose text distance metrics to give poor signal when it comes to evaluating generated code. We found BERT to be much better than BLEU, but the results were still poorer than code-specific distance metrics like TSED. Specifically, the mean BERT score was $98.0\%$, with all 26532 recorded evaluations scoring above $91.0\%$, suggesting potential range restriction and ceiling effects that could negatively impact our experiments were they to use this metric \cite{nunnally1994psychometric}. An example output comparing BERT and TSED is presented in Figure \ref{fig:tsed_vs_bert}. Another consideration were computational costs, where we found BERT to take 20 times longer to compute, even with access to an NVIDIA A100 GPU. Our observations confirm similar conclusions in earlier studies \cite{evtikhiev2023out, song-etal-2024-revisiting}.

\section{Experiments}
We have evaluated 4 LLMs using the pipeline described above:
\begin{itemize}
    \item GPT-4o mini
    \item Claude 3 Haiku
    \item Gemini 2.0 Flash
    \item Llama 3.3 70B
\end{itemize}
To reduce stochasticity of the results, we repeated each request to a LLM 5 times and recorded all responses, and we have also set the temperature of all models to 0. To ensure all requests produce valid code, we added prefixes and postfixes, unaltered and ignored in the analysis, that stressed to the LLM that each request represents a coding task.

It is worth noting that the evaluation pipeline has no dependency on a particular model, and therefore the exact same evaluations can be carried out for any of the modern LLMs. 

\subsection{Overall evaluation}
\begin{figure*}
    \centering
    \subfloat{\includegraphics[width=0.45\textwidth]{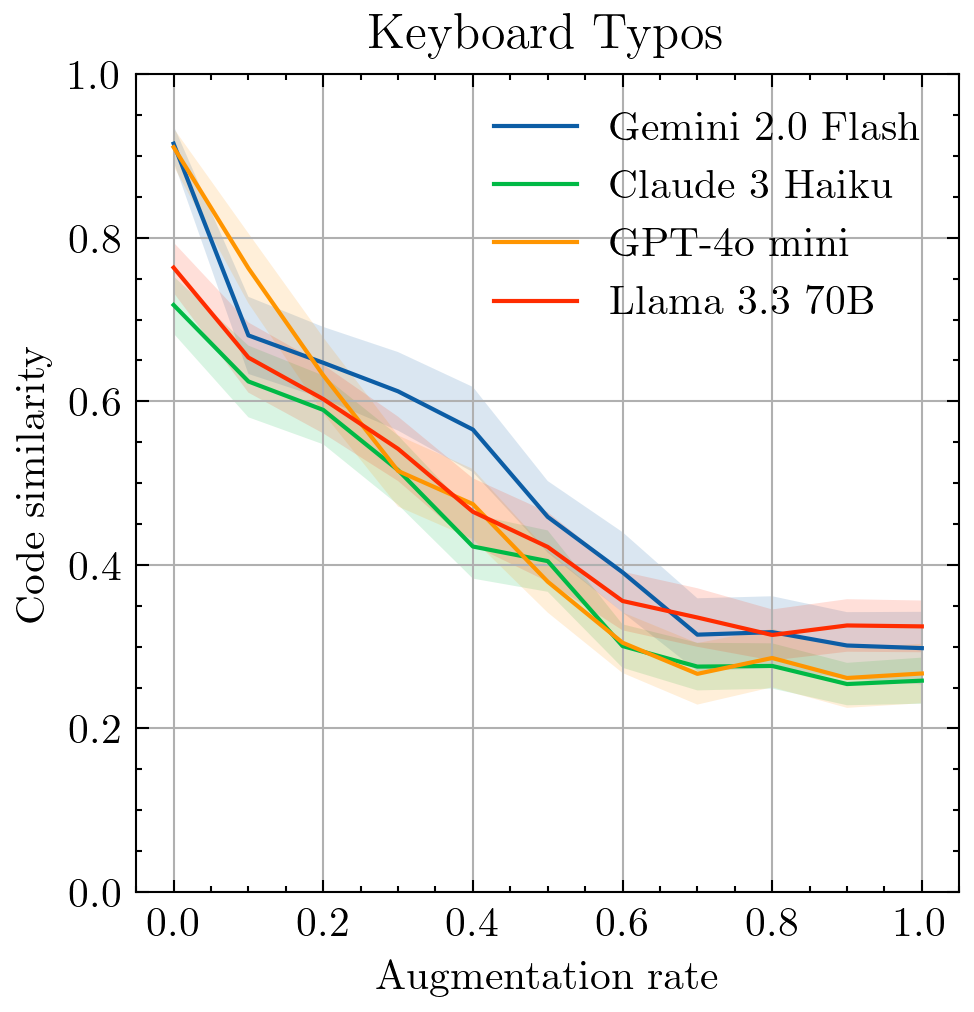}}
    \subfloat{\includegraphics[width=0.45\textwidth]{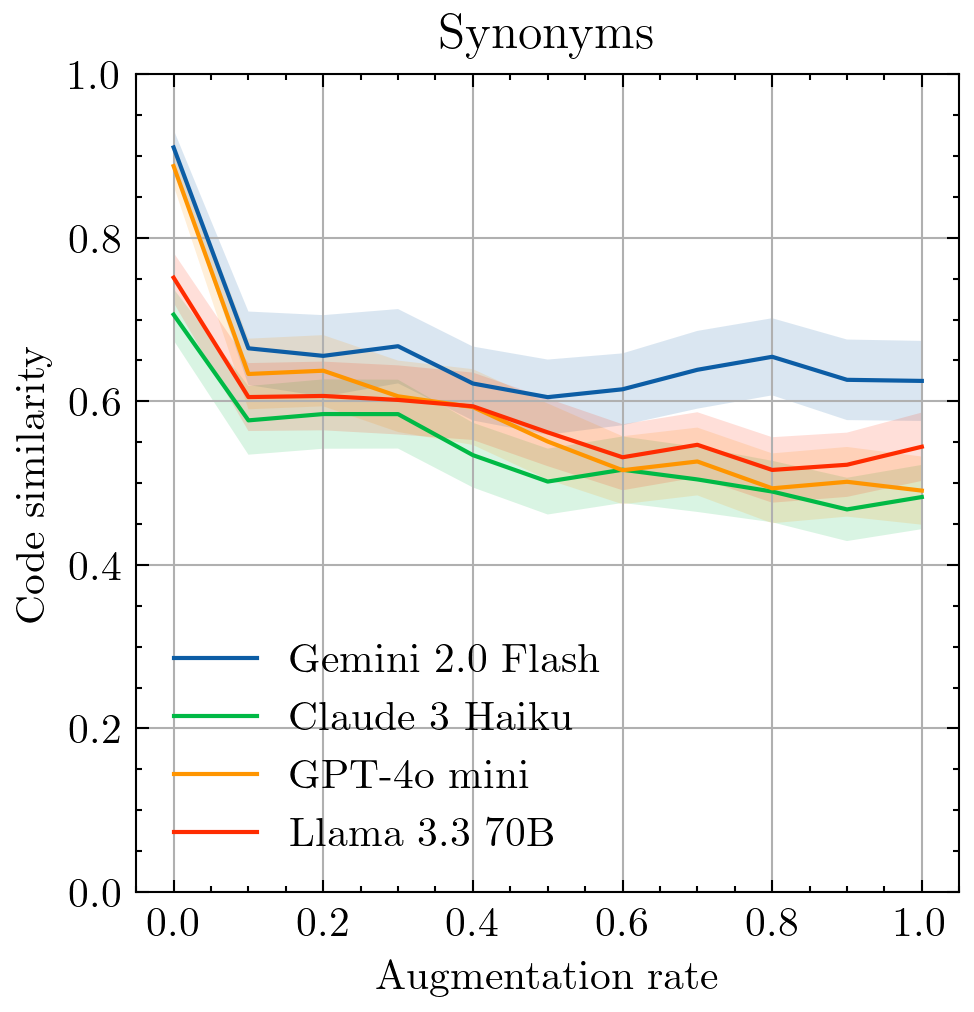}}
    \caption{Overall evaluation results. Solid lines represent mean values and shaded regions the 95\% intervals, calculated from the set of approximately 3400 observations for each rate step. We can confirm that all models exhibit similar sensitivity to prompt augmentations, with Keyboard Typos being a more invasive augmentation method. Gemini 2.0  Flash is the most robust to synonym augmentation, while sensitivity of all models to typos is approximately the same.}
    \label{fig:methods_overall}
\end{figure*}

\begin{figure}
    \centering
    \subfloat{\includegraphics[width=0.45\textwidth]{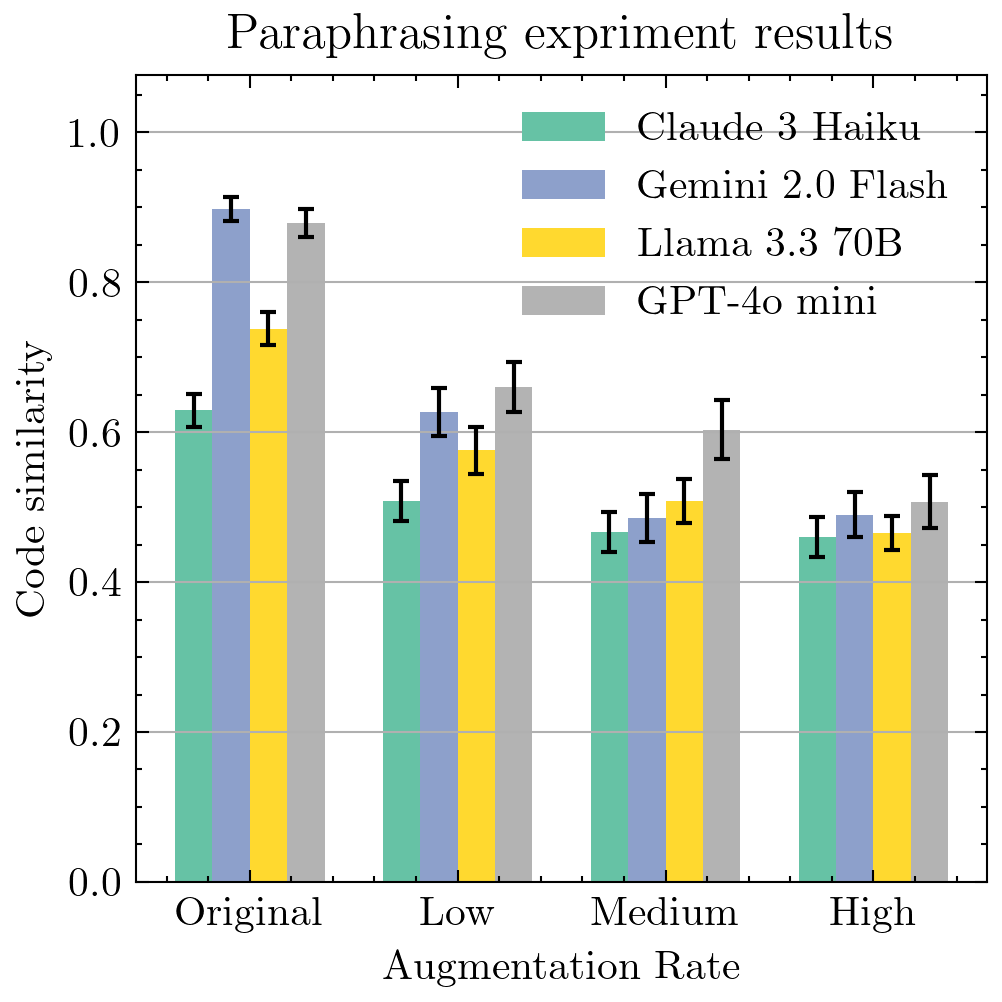}}
    \caption{Results of evaluating the LLM models with paraphrasing augmentation. X-axis shows four different levels of paraphrasing: original (unaltered) as well as low (0.5 - 1.0 BLEU distance), medium (0.2 - 0.5), and high (0.0 - 0.2) from the original. We can see paraphrasing augmentation exhibiting similar trend to synonyms - noticeable drop followed by slow gradual decrease in similarity.}
    \label{fig:paraphrasing_results}
\end{figure}

We begin by presenting overall evaluation results, aggregated over all datasets. These results are shown in Figures \ref{fig:methods_overall} and  \ref{fig:paraphrasing_results}.

All models exhibit similar behavior with respect to \textbf{keyboard typos} augmentation (Figure~\ref{fig:methods_overall}, left), with code similarity dropping rapidly between 0.0 and 0.6 augmentation rate. After reaching a TSED value of approximately 0.3 the decay plateaus. As high augmentation rates with typos essentially lead to unreadable prompts, it is not surprising to see approximately 0.3 TSED score, indicating cardinally different code (compared to the one generated from original prompts).

The observed behavior differs significantly with \textbf{synonyms augmentation} (Figure~\ref{fig:methods_overall}, right), suggesting synonyms being a much weaker prompt augmentation method. After an initial rapid drop at rate 0.1, code outputs vary slightly, eventually dropping just below 0.5 TSED score for Claude 3 Haiku and GPT-4o mini. Gemini 2.0 Flash is the most resilient to this augmentation method, showing higher overall robustness and never dropping below 0.6 in code similarity compared to unaltered prompt's output.

Paraphrasing experiment shows behaviour similar to synonyms, see Figure \ref{fig:paraphrasing_results}. We observe significant drop from the original prompts to low augmentation level, followed by a very gradual decay in code similarity. We can conclude that both synonyms and paraphrasing are much weaker ways of corrupting the prompts, and LLMs are much more robust to such fluctuations in the input compared to typos.

Interestingly, Gemini 2.0 Flash and GPT-4o mini exhibit significantly higher stability for prompts with no augmentation in all experiments, showing 0.9 similarity score with no significant uncertainty. This means these models, with temperature set to 0, output nearly identical codes for the same unaltered prompts. Llama 3.3 and Claude 3 Haiku, on the other hand, exhibit higher instability for unaltered prompts, again with temperature set to 0.

We conducted a Friedman staistical test \cite{friedman1940comparison} to evaluate differences in code similarity across all augmentation rates for this experiment. The test revealed a statistically significant effect of augmentation rate on code similarity ($p < 0.001$ for \textbf{keyboard typos} and \textbf{synonyms}, $p < 0.02$ for \textbf{paraphrasings}), indicating a strong relationship between augmentation rate and LLM code generation sensitivity.

\subsection{Performance between datasets}

\begin{figure*}
    \centering
    \subfloat{\includegraphics[width=0.3\textwidth]{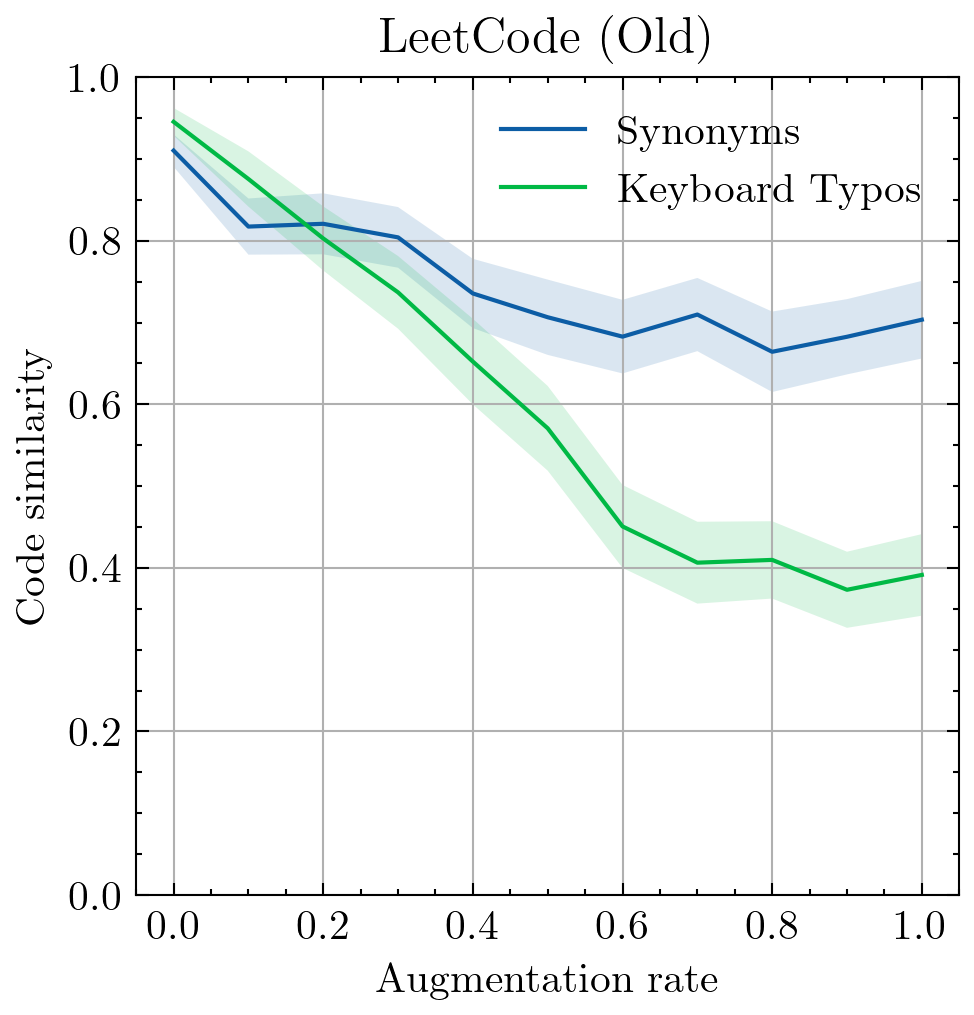}}
    \subfloat{\includegraphics[width=0.3\textwidth]{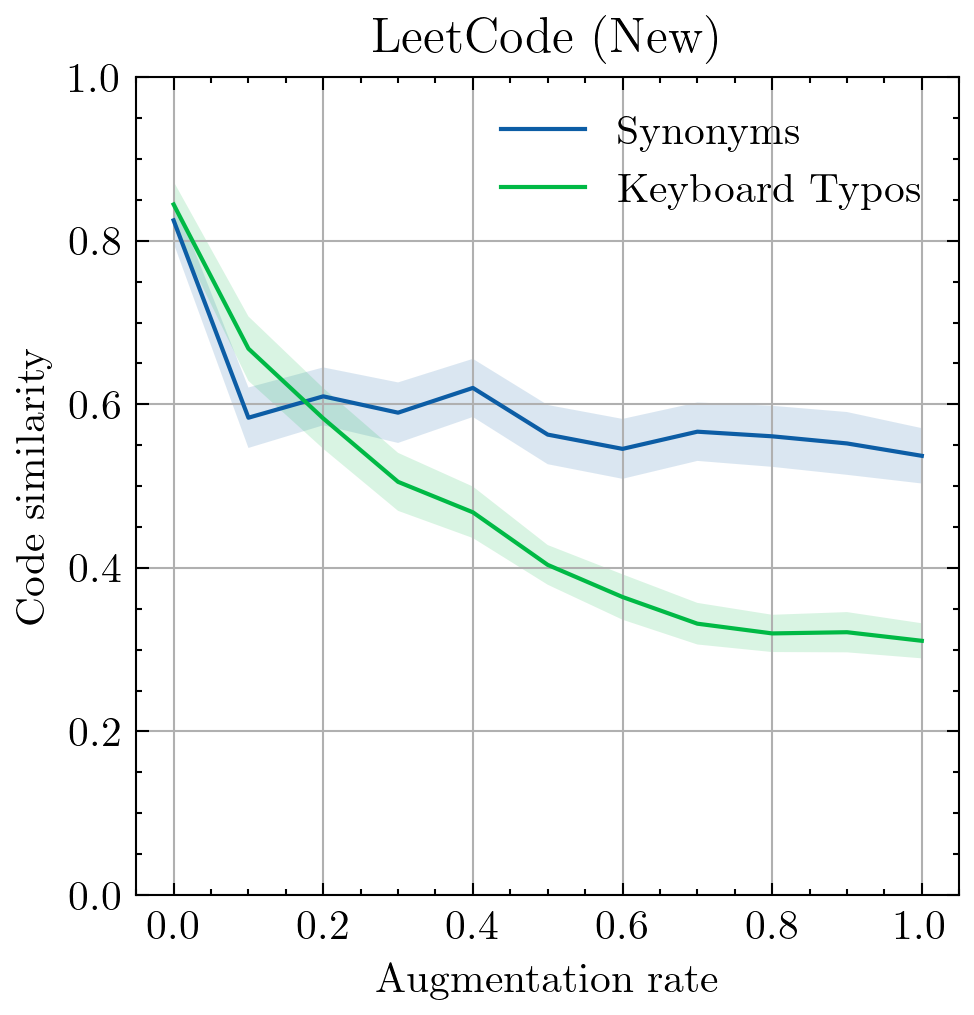}}
    \subfloat{\includegraphics[width=0.3\textwidth]{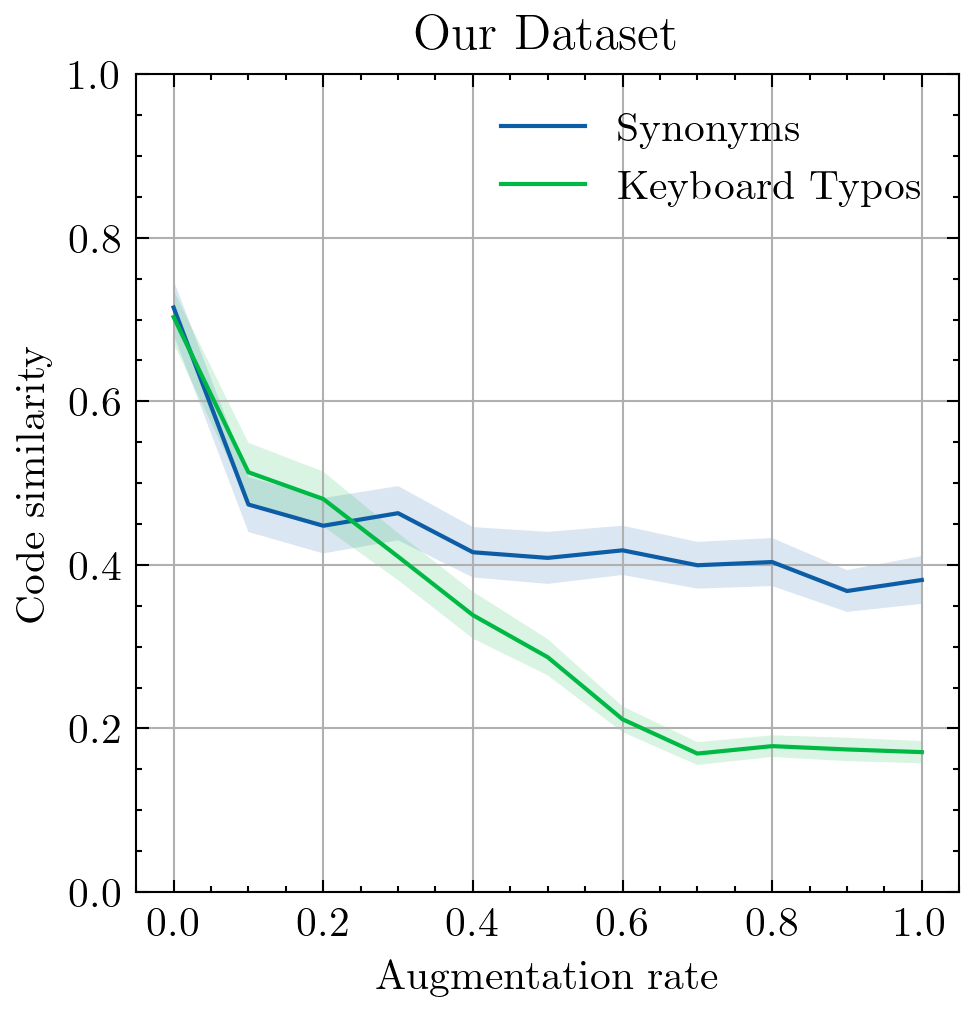}}
    \caption{Evaluation results for three datasets used in this study. Solid lines represent mean values and shaded regions the 95\% intervals. We can see that LLMs show lowest sensitivity to modifications in tasks from LeetCode (Old) dataset, and highest sensitivity sensitivity to tasks from dataset we created from scratch.}
    \label{fig:datasets_overall}
\end{figure*}

Results of evaluating sensitivity on different datasets are shown in Figure~\ref{fig:datasets_overall}

These results illustrate the data contamination phenomenon, previously mentioned in Section \ref{label:data}. Experiments show the lowest sensitivity on \textbf{LeetCode (Old)} dataset across the entire range of augmentation. Because problems in this dataset where most likely included in training of the evaluated models, LLMs can recognise them from comparatively weak signals, showing exceptional robustness even at higher augmentation rates. This issue is discussed in detail by Sendyka et al. \cite{sendyka2025llmperformancecodegeneration}.

This behaviour changes when we consider \textbf{LeetCode (New)} dataset --- code similarity drops lower for both augmentation methods, as these problems were not included in the training data of any of the evaluated models. Nevertheless, newer LeetCode problems still exhibit a lot of traits of standard coding exercises, and therefore all models still exhibit remarkable stability, only dropping below 0.5 similarity score after 50\% of the prompt is altered.

The instability is much more severe on \textbf{Our Dataset}, where problems do not look like a typical coding exercise from LeetCode or competitive programming, and most likely do not correspond directly to any problem found in LLMs' training data. Not only do we observe high variance for even unaltered prompts (only 0.7 similarity between generated code outputs), we can also confirm that code similarity drops below 0.5 TSED after only 10\% of the prompt is modified, suggesting high sensitivity to these modifications.

Taking into account relatively small sizes of our evaluation datasets (not more than 22 tasks per dataset), we have conducted Kruskal-Wallis H-test \cite{kruskal1952use} to test the difference in sensitivity across datasets. The test showed significant ($p < 0.001$ for \textbf{keyboard typos} and $p < 0.02$ for \textbf{synonyms}) difference in the distribution of sensitivity increases across three datasets, supporting the generalizability of our pipeline for a variety of programming tasks.

\section{Related work}
Since LLMs became widely accessible and popular for code generation tasks, the community has focused on their evaluation~\cite{chen2024survey}. This paper extends previous work by proposing an evaluation pipeline for LLM code generation that enables measuring sensitivity to prompt augmentation. This section presents an analysis of the most relevant works in this area.

Initial benchmarks, such as HumanEval \cite{chen2021codex}, SWE-bench \cite{jimenez2023swe}, and Mostly Basic Programming Problems \cite{austin2021program}, enable measuring if generated code meets a programming specification. However, these benchmarks do not include variable prompts and are unfit for the rapidly changing research area, because their datasets and pipelines are static. Some works propose tailored pipelines and frameworks to assess LLMs' code generation capabilities and address the limitations of these initial efforts with more sophisticated and dynamic pipelines. CodeVisionary \cite{wang2025codevisionary} is an agent-based evaluation framework for complex code generation. CodeVisionary relies on a two-step pipeline that includes requirements context distillation and fine-grained scoring. The first step focuses on requirement comprehension and code generation. The second step involves different agents assessing the generated code and reporting the results. CodeArena \cite{du2025codearena} is an online evaluation platform that periodically integrates novel coding tasks in an open repository and offers programming interfaces (APIs) for users' interaction. LiveCodeBench \cite{jain2025livecodebench} is a holistic evaluation benchmark that continuously updates the coding problems from contests across three coding practice platforms (LeetCode, AtCoder, CodeForces). This benchmark implements an evaluation pipeline for code generation, execution, and test output prediction tasks. EvoCodeBench \cite{li2024evocodebench} is a code generation benchmark that dynamically updates code tasks every six months. The benchmark annotates these tasks using a domain taxonomy to enable domain-specific evaluations. The motivation behind continuous updates of coding tasks is to avoid data contamination. This phenomenon threatens evaluations' reproducibility as benchmarking datasets soon become a part of LLMs training data \cite{sendyka2025llmperformancecodegeneration}. Our approach complements these works by including several pipeline stages for prompt augmentation.

Some authors have explored the impact of prompt variations on the code generation results. Murr et al. \cite{murr2311testing} evaluate the performance of LLMs in generating Python programs for coding tasks. They measure how the level of prompt specificity impacts accuracy, time efficiency, and space efficiency of the generated code. The authors propose four levels of specificity, including prompt only, prompt with tests, prompt tests only, and prompt generic tests (i.e., masking test function names). Results show a significant variation in performance across models with different specificity levels. Khojah et al. \cite{khojah2025codeprompeval} evaluate the impact of prompt techniques in LLMs' code generation tasks concerning correctness, similarity, and quality. They create a dataset of 7072 prompts to evaluate prompt techniques such as few-shot, persona, chain-of-thought, function signature, and list of packages. The study shows that prompt techniques, such as few-shot or function signature, positively impact the correctness of the generated code. The persona technique positively impacts the similarity between the generated code and the baseline. Yeo et al. \cite{yeoframework2024} propose a framework to assess the code generation ability of LLMs that integrate with common coding platforms. This framework includes a prompt generation process that creates prompts with different depths. The first level only specifies the problem description, the second level adds constraints, and the third level adds execution examples. Such heterogeneous prompts enable examining their influence on the resulting code. To our knowledge, the three papers above are the closest to our work. These papers report pipelines and results from evaluating the impact of prompt variations in code generation processes. We extend these works by adding augmentation tasks like paraphrasing beyond the proposed specificity levels and prompting techniques, as well as by utilising TSED as a sensitivity measure.

Practitioners have examined the sensitivity of LLMs in domains other than code. The PromptSET benchmark adopts a methodology of slight modifications to prompts, but it evaluates general tasks \cite{razavi2025benchmarking}. Gulati et al. \cite{gulati2024putnam} evaluated LLMs on mathematical problems from the Putnam Competition with variable names and constant alterations. They found that performance drops significantly on altered inputs compared to the originals. Their evaluation technique closely resembles our experiments. Hu et al. \cite{hu2024prompt} performed similar experiments on different datasets, including IMDB WikiData, Opendatasoft. Sclar et al. \cite{sclar2023quantifying} investigated the importance of prompt formatting for reporting LLM performance. Chan et al. \cite{chan2025effectiveness} use metamorphic testing to validate the stability of code-generating LLMs. This paper evaluates LLMs' sensitivity to prompt perturbations under a completely different setting, namely, their primary use case is a private software house downloading and pretraining a code generation model to create large software projects, meaning their primary focus is on private models. We propose considering a scenario of the general public interacting with widespread and publicly accessible LLM APIs for relatively simple coding tasks. Together, these efforts provide a complementary coverage of LLMs' code generation sensitivity in different use cases.

\section{Threats to validity}
In this section we discuss main threats to validity of our study.

\textbf{Internal validity.} Large Language Models are normally trained to encourage variability of their output. While temperature is considered to be the main parameter that gauges how random the output of a LLM is, even setting temperature to 0 does not guarantee fully deterministic results. To alleviate this effect we have generated large quantities of raw LLM responses to aggregate over (reaching approximately 3400 datapoints per single augmentation rate step), thus smoothing over occasional anomalous outputs.

As mentioned before, TSED measures structural similarity, not functional equivalence, creating potential construct validity issues. However, this aligns with our goal of measuring LLM sensitivity and output consistency. Structural variability poses real challenges for users who must understand, debug, and maintain different implementations. We recommend complementing TSED with functional testing when correctness is critical.

\textbf{External validity.} The world of LLMs develops at a rapid pace, with new models, or new versions of existing models, being released at frequently\footnote{For example, this Reddit post lists 12 releases in March-May 2024: \url{https://www.reddit.com/r/LocalLLaMA/comments/1cn9sxa/timeline_of_recent_major_llm_releases_past_2/}}. With the constantly moving frontier and relatively slow pace of academic publication, any study on LLMs is bound to be missing most recent releases by the time it is published. Being aware of this risk, we designed our pipeline to be completely agnostic of specific models or API implementations. We have also used multiple LLM providers in all our experiments to show how our method can be applied to different models.

\section{Conclusion}
Using LLMs to generate code is a proliferating use case. Different users can describe the same code requirement in their prompts in slightly different words, leading to unintentional variation in generated code, thus motivating the need to understand how sensitive LLMs are to fluctuations in prompts. This understanding can help build trust in the code generation process, inform industry providers that create code generation pipelines, and help uncover gaps to motivate further research into LLMs. We have proposed a novel pipeline for evaluating sensitivity of code generation with LLMs to changes in prompts and have shown preliminary evidence that even small fluctuations in prompts such as typos, synonyms, and paraphrasing can impact the generated code. 

Concretely, we described an evaluation procedure that uses keyboard typos, synonyms, and paraphrasing to increasingly alter the code task prompt, and uses syntax-tree based code similarity metric to measure changes in the code output. After evaluating specific versions of 4 popular LLMs we find all models to exhibit similar behaviour, showing rapid decay of code similarity for typos augmentation, and being more robust to augmentation with synonyms and paraphrasing. We also illustrate an issue known as data contamination, argue for use of more recent and original tasks to evaluate LLM code generation, and provide a new dataset of open ended tasks.

There can be multiple ways to extend ideas and methods presented in this paper. So far we have only considered one-step interactions with LLMs, while quite often these interactions resemble a dialog. It is therefore reasonable to extend our methods to model such dialogs, using metrics such as the number of information exchanges to measure quality. Similarly, our evaluation can be applied to multi-step pipelines, quantifying uncertainty propagation in such settings. Furthermore, complementing TSED with functional testing would provide a more complete measure of structural and functional variablity of the generated code. Finally, it would be valuable to conduct similar experiments involving participants with diverse levels of experience in software engineering, or using augmentation methods that model differences in problem decomposition, mental models, and conceptual framing across users with different backgrounds.

We hope to have provided a deeper insight of the nuanced ways in which prompt variation impacts code generation with LLMs. Ultimately, we aim to inspire future work that enhances the robustness, reliability, and trust in LLM-based code generation systems.

\begin{acks}
This project is supported by funding from the Accelerate Programme for Scientific Discovery, made possible by a donation from Schmidt Sciences.
\end{acks}

\bibliographystyle{ACM-Reference-Format}
\bibliography{references}

@article{zhang2023unifying,
  title={Unifying the perspectives of nlp and software engineering: A survey on language models for code},
  author={Zhang, Ziyin and Chen, Chaoyu and Liu, Bingchang and Liao, Cong and Gong, Zi and Yu, Hang and Li, Jianguo and Wang, Rui},
  journal={arXiv preprint arXiv:2311.07989},
  year={2023}
}

@article{robinson2024requirements,
  title={Requirements are all you need: The final frontier for end-user software engineering},
  author={Robinson, Diana and Cabrera, Christian and Gordon, Andrew D and Lawrence, Neil D and Mennen, Lars},
  journal={ACM Transactions on Software Engineering and Methodology},
  volume={34},
  number={5},
  pages={1--22},
  year={2025},
  publisher={ACM New York, NY}
}

@ARTICLE{Heinonen2023-ac,
  title     = "Synthesizing research on programmers’ mental models of programs,
               tasks and concepts — A systematic literature review",
  author    = "Heinonen, Ava and Lehtelä, Bettina and Hellas, Arto and
               Fagerholm, Fabian",
  journal   = "Inf. Softw. Technol.",
  publisher = "Elsevier BV",
  volume    =  164,
  number    =  107300,
  pages     =  107300,
  month     =  dec,
  year      =  2023,
  language  = "en"
}

@ARTICLE{Miaskiewicz2011-qy,
  title     = "Personas and user-centered design: How can personas benefit
               product design processes?",
  author    = "Miaskiewicz, Tomasz and Kozar, Kenneth A",
  journal   = "Des. Stud.",
  publisher = "Elsevier BV",
  volume    =  32,
  number    =  5,
  pages     = "417--430",
  month     =  sep,
  year      =  2011,
  language  = "en"
}

@INPROCEEDINGS{El-Assady2019-rg,
  title     = "Lingvis.Io - A linguistic visual analytics framework",
  author    = "El-Assady, Mennatallah and Jentner, Wolfgang and Sperrle, Fabian
               and Sevastjanova, Rita and Hautli-Janisz, Annette and Butt,
               Miriam and Keim, Daniel",
  booktitle = "Proceedings of the 57th Annual Meeting of the Association for
               Computational Linguistics: System Demonstrations",
  publisher = "Association for Computational Linguistics",
  address   = "Stroudsburg, PA, USA",
  pages     = "13--18",
  year      =  2019
}

@ARTICLE{Blei2003-zm,
  title     = "Latent dirichlet allocation",
  author    = "Blei, David M and Ng, Andrew Y and Jordan, Michael I",
  journal   = "J. Mach. Learn. Res.",
  publisher = "JMLR.org",
  volume    =  3,
  number    = "null",
  pages     = "993--1022",
  month     =  mar,
  year      =  2003,
  language  = "en"
}

@ARTICLE{Jung2025-om,
  title     = "{PersonaCraft}: Leveraging language models for data-driven
               persona development",
  author    = "Jung, Soon-Gyo and Salminen, Joni and Aldous, Kholoud Khalil and
               Jansen, Bernard J",
  journal   = "Int. J. Hum. Comput. Stud.",
  publisher = "Elsevier BV",
  volume    =  197,
  number    =  103445,
  pages     =  103445,
  month     =  mar,
  year      =  2025,
  language  = "en"
}

@ARTICLE{Ehsani2025-cy,
  title         = "Towards detecting prompt knowledge gaps for improved
                   {LLM}-guided issue resolution",
  author        = "Ehsani, Ramtin and Pathak, Sakshi and Chatterjee, Preetha",
  journal       = "arXiv [cs.SE]",
  month         =  jan,
  year          =  2025,
  archivePrefix = "arXiv",
  primaryClass  = "cs.SE"
}

@ARTICLE{Liang2024-eh,
  title         = "Prompts are programs too! Understanding how developers build
                   software containing prompts",
  author        = "Liang, Jenny T and Lin, Melissa and Rao, Nikitha and Myers,
                   Brad A",
  journal       = "arXiv [cs.SE]",
  month         =  sep,
  year          =  2024,
  archivePrefix = "arXiv",
  primaryClass  = "cs.SE"
}

@ARTICLE{Guo2024-hd,
  title         = "{M}-ped: Multi-prompt ensemble decoding for Large Language
                   Models",
  author        = "Guo, Jiaxin and Wei, Daimeng and Luo, Yuanchang and Tao,
                   Shimin and Shang, Hengchao and Li, Zongyao and Li, Shaojun
                   and Yang, Jinlong and Wu, Zhanglin and Rao, Zhiqiang and
                   Yang, Hao",
  journal       = "arXiv [cs.CL]",
  month         =  dec,
  year          =  2024,
  archivePrefix = "arXiv",
  primaryClass  = "cs.CL"
}

@MISC{Neumann2021-ju,
  title        = "Teaching Computational Thinking",
  author       = "Neumann, Maureen D and Dion, Lisa and Snapp, Robert",
  booktitle    = "MIT Press",
  publisher    = "The MIT Press, Massachusetts Institute of Technology",
  month        =  dec,
  year         =  2021,
  howpublished = "\url{https://mitpress.mit.edu/9780262045056/teaching-computational-thinking/}",
  note         = "Accessed: 2025-5-15",
  language     = "en"
}

@INPROCEEDINGS{Poitras2024-wa,
  title     = "Generative {AI} in introductory programming instruction:
               Examining the assistance dilemma with {LLM}-based code generators",
  author    = "Poitras, Eric and Crane, Brent and Siegel, Angela",
  booktitle = "Proceedings of the 2024 on ACM Virtual Global Computing Education
               Conference V. 1",
  publisher = "ACM",
  address   = "New York, NY, USA",
  pages     = "186--192",
  month     =  dec,
  year      =  2024
}

@INPROCEEDINGS{Vadaparty2024-pi,
  title     = "{CS1}-{LLM}: Integrating {LLMs} into {CS1} Instruction",
  author    = "Vadaparty, Annapurna and Zingaro, Daniel and Smith, IV, David H
               and Padala, Mounika and Alvarado, Christine and Gorson Benario,
               Jamie and Porter, Leo",
  booktitle = "Proceedings of the 2024 on Innovation and Technology in Computer
               Science Education V. 1",
  publisher = "ACM",
  address   = "New York, NY, USA",
  pages     = "297--303",
  month     =  jul,
  year      =  2024
}

@ARTICLE{Subramonyam2024-ow,
  title         = "Prototyping with prompts: Emerging approaches and challenges
                   in generative {AI} design for collaborative software teams",
  author        = "Subramonyam, Hari and Thakkar, Divy and Ku, Andrew and
                   Dieber, Jürgen and Sinha, Anoop",
  journal       = "arXiv [cs.HC]",
  month         =  feb,
  year          =  2024,
  archivePrefix = "arXiv",
  primaryClass  = "cs.HC"
}

@ARTICLE{Wing2008-re,
  title     = "Computational thinking and thinking about computing",
  author    = "Wing, Jeannette M",
  journal   = "Philos. Trans. A Math. Phys. Eng. Sci.",
  publisher = "The Royal Society",
  volume    =  366,
  number    =  1881,
  pages     = "3717--3725",
  month     =  oct,
  year      =  2008,
  language  = "en"
}

@ARTICLE{Chen2024-nd,
author = {Chen, Junkai and Zhenhao, Li and Xing, Hu and Xin, Xia},
title = {NLPerturbator: Studying the Robustness of Code LLMs to Natural Language Variations},
year = {2025},
publisher = {Association for Computing Machinery},
address = {New York, NY, USA},
issn = {1049-331X},
url = {https://doi.org/10.1145/3745764},
doi = {10.1145/3745764},
note = {Just Accepted},
journal = {ACM Trans. Softw. Eng. Methodol.},
month = jul
}

@inproceedings{gulati2024putnam,
  title={{Putnam-AXIOM}: A functional and static benchmark for measuring higher level mathematical reasoning},
  author={Gulati, Aryan and Miranda, Brando and Chen, Eric and Xia, Emily and Fronsdal, Kai and de Moraes Dumont, Bruno and Koyejo, Sanmi},
  booktitle={The 4th Workshop on Mathematical Reasoning and AI at NeurIPS'24},
  year={2024}
}

@article{chan2025effectiveness,
  title={Effectiveness of symmetric metamorphic relations on validating the stability of code generation {LLM}},
  author={Chan, Pak Yuen Patrick and Keung, Jacky and Yang, Zhen},
  journal={Journal of Systems and Software},
  volume={222},
  pages={112330},
  year={2025},
  publisher={Elsevier}
}

@article{chen2021codex,
  title={Evaluating Large Language Models Trained on Code},
  author={Mark Chen and Jerry Tworek and Heewoo Jun and Qiming Yuan and Henrique Ponde de Oliveira Pinto and Jared Kaplan and Harri Edwards and Yuri Burda and Nicholas Joseph and Greg Brockman and Alex Ray and Raul Puri and Gretchen Krueger and Michael Petrov and Heidy Khlaaf and Girish Sastry and Pamela Mishkin and Brooke Chan and Scott Gray and Nick Ryder and Mikhail Pavlov and Alethea Power and Lukasz Kaiser and Mohammad Bavarian and Clemens Winter and Philippe Tillet and Felipe Petroski Such and Dave Cummings and Matthias Plappert and Fotios Chantzis and Elizabeth Barnes and Ariel Herbert-Voss and William Hebgen Guss and Alex Nichol and Alex Paino and Nikolas Tezak and Jie Tang and Igor Babuschkin and Suchir Balaji and Shantanu Jain and William Saunders and Christopher Hesse and Andrew N. Carr and Jan Leike and Josh Achiam and Vedant Misra and Evan Morikawa and Alec Radford and Matthew Knight and Miles Brundage and Mira Murati and Katie Mayer and Peter Welinder and Bob McGrew and Dario Amodei and Sam McCandlish and Ilya Sutskever and Wojciech Zaremba},
  year={2021},
  eprint={2107.03374},
  archivePrefix={arXiv},
  primaryClass={cs.LG}
}

@inproceedings{sclar2023quantifying,
  title={Quantifying Language Models' Sensitivity to Spurious Features in Prompt Design or: How I learned to start worrying about prompt formatting},
  author={Sclar, Melanie and Choi, Yejin and Tsvetkov, Yulia and Suhr, Alane},
  booktitle={The Twelfth International Conference on Learning Representations},
  year={2023}
}

@inproceedings{jimenez2023swe,
  title={{SWE-bench}: Can Language Models Resolve Real-world {GitHub} Issues?},
  author={Jimenez, Carlos E and Yang, John and Wettig, Alexander and Yao, Shunyu and Pei, Kexin and Press, Ofir and Narasimhan, Karthik R},
  booktitle={The Twelfth International Conference on Learning Representations},
  year={2023}
}

@article{austin2021program,
  title={Program synthesis with large language models},
  author={Austin, Jacob and Odena, Augustus and Nye, Maxwell and Bosma, Maarten and Michalewski, Henryk and Dohan, David and Jiang, Ellen and Cai, Carrie and Terry, Michael and Le, Quoc and others},
  journal={arXiv preprint arXiv:2108.07732},
  year={2021}
}

@inproceedings{hu2024prompt,
  title={Prompt perturbation in retrieval-augmented generation based large language models},
  author={Hu, Zhibo and Wang, Chen and Shu, Yanfeng and Paik, Hye-Young and Zhu, Liming},
  booktitle={Proceedings of the 30th ACM SIGKDD Conference on Knowledge Discovery and Data Mining},
  pages={1119--1130},
  year={2024}
}

@article{evtikhiev2023out,
  title={Out of the {Bleu}: how should we assess quality of the code generation models?},
  author={Evtikhiev, Mikhail and Bogomolov, Egor and Sokolov, Yaroslav and Bryksin, Timofey},
  journal={Journal of Systems and Software},
  volume={203},
  pages={111741},
  year={2023},
  publisher={Elsevier}
}

@inproceedings{razavi2025benchmarking,
  title={Benchmarking prompt sensitivity in large language models},
  author={Razavi, Amirhossein and Soltangheis, Mina and Arabzadeh, Negar and Salamat, Sara and Zihayat, Morteza and Bagheri, Ebrahim},
  booktitle={European Conference on Information Retrieval},
  pages={303--313},
  year={2025},
  organization={Springer}
}

@misc{ma2019nlpaug,
  title={{NLP} Augmentation},
  author={Edward Ma},
  howpublished={https://github.com/makcedward/nlpaug},
  year={2019}
}

@article{miller1995wordnet,
  title={{WordNet}: a lexical database for {E}nglish},
  author={Miller, George A},
  journal={Communications of the ACM},
  volume={38},
  number={11},
  pages={39--41},
  year={1995},
  publisher={ACM New York, NY, USA}
}

@inproceedings{song-etal-2024-revisiting,
    title = "Revisiting Code Similarity Evaluation with Abstract Syntax Tree Edit Distance",
    author = "Song, Yewei  and
      Lothritz, Cedric  and
      Tang, Xunzhu  and
      Bissyand{\'e}, Tegawend{\'e}  and
      Klein, Jacques",
    editor = "Ku, Lun-Wei  and
      Martins, Andre  and
      Srikumar, Vivek",
    booktitle = "Proceedings of the 62nd Annual Meeting of the Association for Computational Linguistics (Volume 2: Short Papers)",
    month = aug,
    year = "2024",
    address = "Bangkok, Thailand",
    publisher = "Association for Computational Linguistics",
    url = "https://aclanthology.org/2024.acl-short.3/",
    doi = "10.18653/v1/2024.acl-short.3",
    pages = "38--46"
}

@inproceedings{zhang2020bertscore,
  title={{BERTScore}: Evaluating Text Generation with {BERT}},
  author={Zhang, Tianyi and Kishore, Varsha and Wu, Felix and Weinberger, Kilian Q and Artzi, Yoav},
  booktitle={International Conference on Learning Representations},
  year={2020}
}

@inproceedings{post-2018-call,
    title = "A Call for Clarity in Reporting {BLEU} Scores",
    author = "Post, Matt",
    editor = "Bojar, Ond{\v{r}}ej  and
      Chatterjee, Rajen  and
      Federmann, Christian  and
      Fishel, Mark  and
      Graham, Yvette  and
      Haddow, Barry  and
      Huck, Matthias  and
      Yepes, Antonio Jimeno  and
      Koehn, Philipp  and
      Monz, Christof  and
      Negri, Matteo  and
      N{\'e}v{\'e}ol, Aur{\'e}lie  and
      Neves, Mariana  and
      Post, Matt  and
      Specia, Lucia  and
      Turchi, Marco  and
      Verspoor, Karin",
    booktitle = "Proceedings of the Third Conference on Machine Translation: Research Papers",
    month = oct,
    year = "2018",
    address = "Brussels, Belgium",
    publisher = "Association for Computational Linguistics",
    url = "https://aclanthology.org/W18-6319/",
    doi = "10.18653/v1/W18-6319",
    pages = "186--191"
}

@misc{sendyka2025llmperformancecodegeneration,
      title={LLM Performance for Code Generation on Noisy Tasks}, 
      author={Radzim Sendyka and Christian Cabrera and Andrei Paleyes and Diana Robinson and Neil Lawrence},
      year={2025},
      eprint={2505.23598},
      archivePrefix={arXiv},
      primaryClass={cs.LG},
      url={https://arxiv.org/abs/2505.23598}, 
}

@article{friedman1940comparison,
  title={A comparison of alternative tests of significance for the problem of m rankings},
  author={Friedman, Milton},
  journal={The annals of mathematical statistics},
  volume={11},
  number={1},
  pages={86--92},
  year={1940},
  publisher={JSTOR}
}

@article{kruskal1952use,
  title={Use of ranks in one-criterion variance analysis},
  author={Kruskal, William H and Wallis, W Allen},
  journal={Journal of the American statistical Association},
  volume={47},
  number={260},
  pages={583--621},
  year={1952},
  publisher={Taylor \& Francis}
}

@article{chen2024survey,
  title={A Survey on Evaluating Large Language Models in Code Generation Tasks},
  author={Chen, Liguo and Guo, Qi and Jia, Hongrui and Zeng, Zhengran and Wang, Xin and Xu, Yijiang and Wu, Jian and Wang, Yidong and Gao, Qing and Wang, Jindong and others},
  journal={arXiv e-prints},
  pages={arXiv--2408},
  year={2024}
}

@article{du2025codearena,
  publtype={informal},
  author={Mingzhe Du and Anh Tuan Luu and Bin Ji and Xiaobao Wu and Dong Huang and Terry Yue Zhuo and Qian Liu and See-Kiong Ng},
  title={CodeArena: A Collective Evaluation Platform for LLM Code Generation},
  year={2025},
  month={March},
  cdate={1740787200000},
  journal={CoRR},
  volume={abs/2503.01295},
  url={https://doi.org/10.48550/arXiv.2503.01295}
}

@inproceedings{jain2025livecodebench,
title={LiveCodeBench: Holistic and Contamination Free Evaluation of Large Language Models for Code},
author={Naman Jain and King Han and Alex Gu and Wen-Ding Li and Fanjia Yan and Tianjun Zhang and Sida Wang and Armando Solar-Lezama and Koushik Sen and Ion Stoica},
booktitle={The Thirteenth International Conference on Learning Representations},
year={2025},
url={https://openreview.net/forum?id=chfJJYC3iL}
}

@article{yeoframework2024,
author = {Yeo, Sangyeop and Ma, Yu-Seung and Kim, Sang Cheol and Jun, Hyungkook and Kim, Taeho},
title = {Framework for evaluating code generation ability of large language models},
journal = {ETRI Journal},
volume = {46},
number = {1},
pages = {106-117},
doi = {https://doi.org/10.4218/etrij.2023-0357},
url = {https://onlinelibrary.wiley.com/doi/abs/10.4218/etrij.2023-0357},
year = {2024}
}

@article{wang2025codevisionary,
  title={CodeVisionary: An Agent-based Framework for Evaluating Large Language Models in Code Generation},
  author={Wang, Xinchen and Gao, Pengfei and Peng, Chao and Hu, Ruida and Gao, Cuiyun},
  journal={arXiv e-prints},
  pages={arXiv--2504},
  year={2025}
}

@inproceedings{li2024evocodebench,
 author = {Li, Jia and Li, Ge and Zhang, Xuanming and Zhao, Yunfei and Dong, Yihong and Jin, Zhi and Li, Binhua and Huang, Fei and Li, Yongbin},
 booktitle = {Advances in Neural Information Processing Systems},
 editor = {A. Globerson and L. Mackey and D. Belgrave and A. Fan and U. Paquet and J. Tomczak and C. Zhang},
 pages = {57619--57641},
 publisher = {Curran Associates, Inc.},
 title = {EvoCodeBench: An Evolving Code Generation Benchmark with Domain-Specific Evaluations},
 volume = {37},
 year = {2024}
}

@article{murr2311testing,
  title={Testing LLMS on code generation with varying levels of prompt specificity},
  author={Murr, Lincoln and Grainger, Morgan and Gao, David},
  journal={URL https://arxiv. org/abs/2311.07599},
  year={2023}
}

@ARTICLE{khojah2025codeprompeval,
  author={Khojah, Ranim and de Oliveira Neto, Francisco Gomes and Mohamad, Mazen and Leitner, Philipp},
  journal={IEEE Transactions on Software Engineering}, 
  title={The Impact of Prompt Programming on Function-Level Code Generation}, 
  year={2025},
  volume={51},
  number={8},
  pages={2381-2395},
  doi={10.1109/TSE.2025.3587794}
}

@book{nunnally1994psychometric,
  title     = {Psychometric Theory},
  author    = {Nunnally, Jum C. and Bernstein, Ira H.},
  edition   = {3rd},
  year      = {1994},
  publisher = {McGraw-Hill},
  address   = {New York},
  pages     = {xxiv + 752}
}

\clearpage
\appendix

\section{Appendix: Persona evaluation}\label{section:persona-eval}
Our synthetic experiments show that LLMs can indeed be sensitive to variations in code generation tasks. While these experiments can be useful for benchmarking purposes, the augmentation methods we utilised do not fully resemble the kind of difference we might expect to see between two humans expressing the same task as an LLM prompt.

To test our ideas that the description of a programming task will vary depending on the programming experience and background of the person providing it, we have defined personas with different software engineering experience. Using these personas, we have each generate a prompt for a particular coding task and then run the prompts in ChatGPT and Claude. We investigate whether there are qualitative  differences in both the prompts and code generated by the different personas on three tasks.

\subsection{Defining personas}

In human-computer interaction, researchers sometimes create personas that are based on a summary of characteristics of potential users of the system of interest. User personas provide a description of a person or summary of a set of people, which can include their background, description of the knowledge they have or personality, if relevant \cite{Miaskiewicz2011-qy}. Recently, with the widespread use of LLMs, there is potential for LLMs to take on such personas and to use them in studies, including as proxies for real humans \cite{Jung2025-om}.

In this section, we created persona descriptions then prompted an LLM to adopt them and generate a prompt, in order to test our hypothesis that background and knowledge of different people will affect how they will decompose a problem. We then observed the effects this had on the generated code. 

When writing the persona descriptions, we focused on varying experience with programming and software development to test this aspect of background difference in how they explain a task and decompose a problem. For the descriptions of background and education, we searched profiles on LinkedIn and job descriptions in order to look at skills and knowledge reflective of these roles. We used four personas which were a junior software engineer with a computer science degree and internship experience, a principal software engineer with significant industry experience, an english teacher who had never used a programming language and was unfamiliar with computing concepts but able to use common software like Microsoft Word and Excel, and an astrophysicist who used programming languages like Python in the context of research but was not a software engineer. We provide descriptions in the linked GitHub repo. We used all male personas in order to isolate whether programming experience influences prompt development and the resulting code without influence from gender bias in the training set of the LLM. This is a topic that future work should explore.

The setup of the experiment was to instruct the LLM to adopt each persona in turn and to paraphrase a provided prompt, using their background and assumed knowledge to inform the language they use to describe the task and to add any details that they consider relevant. For the tasks, we selected three from the dataset we developed (see details in Section \ref{subsection:data}, Our Dataset). For this experiment, we underspecified the prompt for the LLM persona to paraphrase, in order to test how each persona broke down the task and explained it and to see if the length of the descriptions and the information they added varied with persona descriptions. For example, in the controlled evaluation in Section \ref{subsection:data}, the prompt was ``Write a Calculator class. It shall contain common arithmetic operations, such as addition or multiplication, but also more advanced operations, such as logarithm (of variable bases), factorial, trigonometry, roots, exponents.'' whereas in the personas task, we made it ``Write code to make a calculator'' to better simulate the starting point that someone would have before breaking down the problem and explaining it in their own words.

In order to test the stability of the personas in generating prompts, we ran five iterations for each persona for each of the three tasks. We then randomly sampled one of these prompts for each persona and input it into the user interface of ChatGPT and Claude to generate the code.

\subsection{Evaluation}

We analysed the prompts as well as the generated code for each persona to examine whether and how the personas affected the generated code.

For the prompt analysis, we used a linguistic visual analytics framework from El-Assady et al. \cite{El-Assady2019-rg} which provides a pipeline for modelling entities (specific pieces of text) within a corpus, extracting features and applying latent dirichlet allocation (LDA) for topic modelling \cite{Blei2003-zm} and then visualising the number of entities (specific pieces of text) which are used in close proximity within sentences between speakers. This provides a measure of how similar different speakers are in their style which we use to analyse the prompts provided by the different personas. 

The tasks that we generated prompts for were: write code to make a calculator, write code for a database for an online bike shop, and write code to generate accounting reports. The code generated by each persona on the three tasks and two models (ChatGPT and Claude) was evaluated observationally for trends. 

\subsection{Results}
\begin{figure*}
    \centering
    \subfloat{\includegraphics[width=0.45\textwidth]{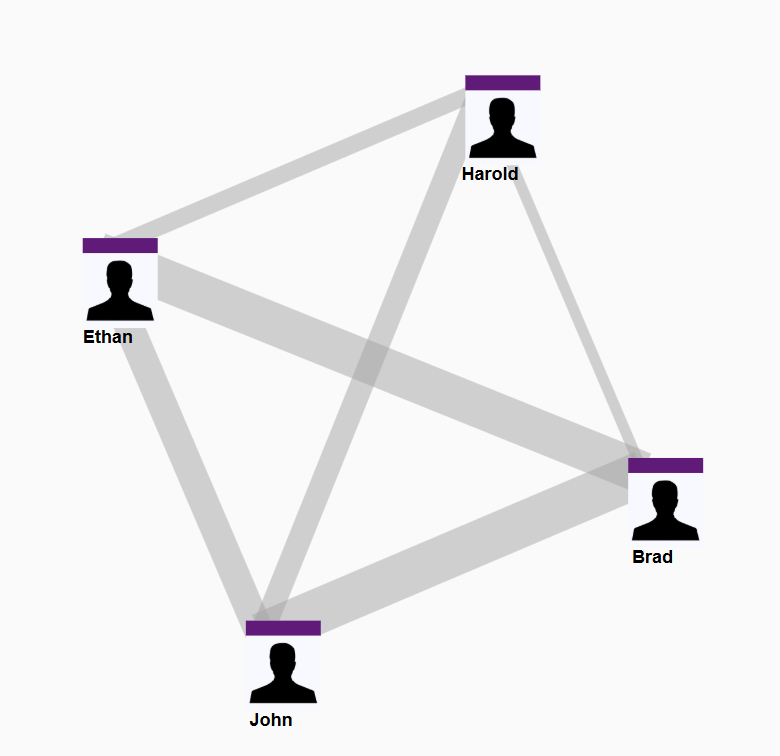}}
    \subfloat{\includegraphics[width=0.45\textwidth]{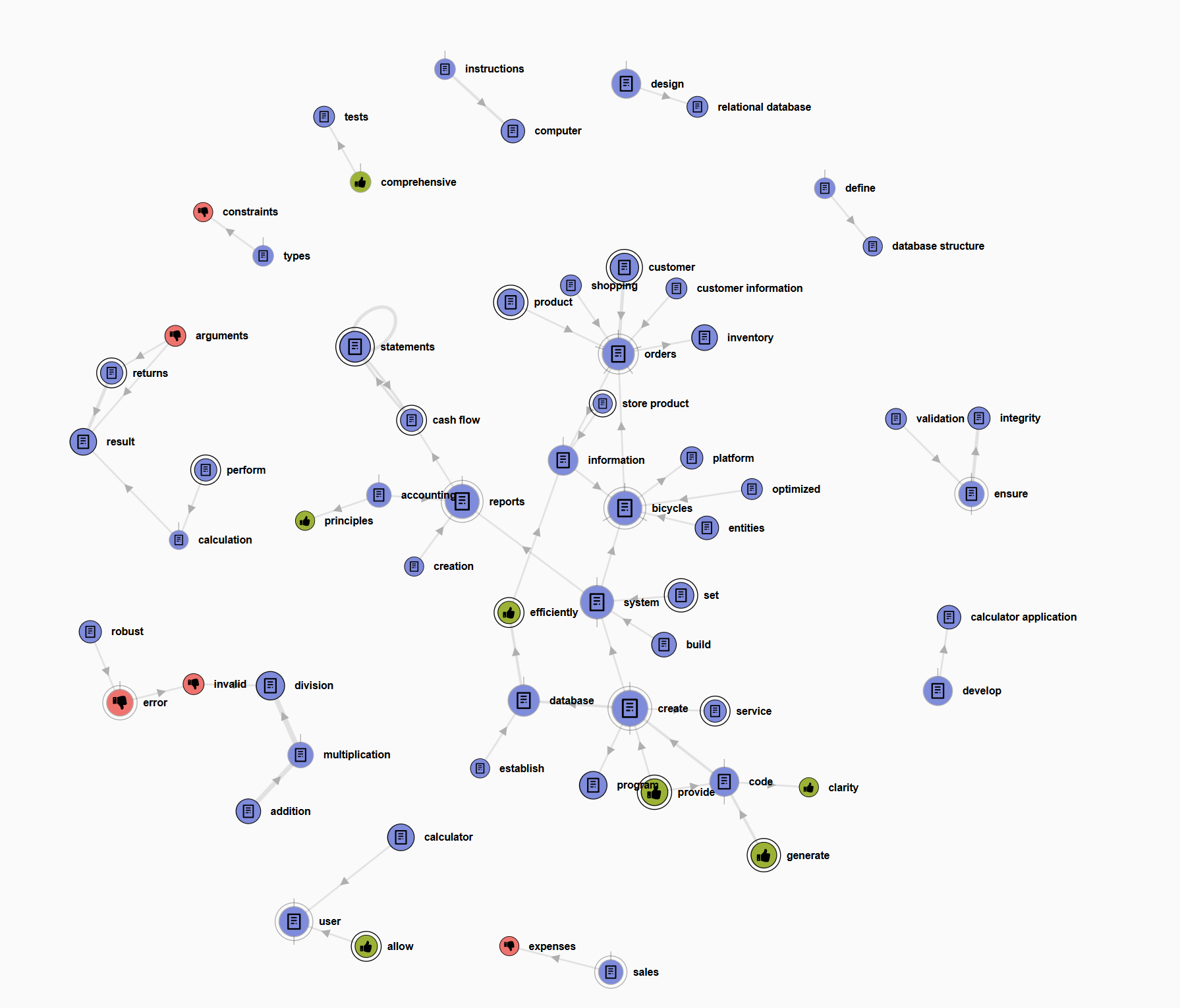}}
    \caption{Visualisation of the number of entities (specific pieces of text) which are used in close proximity within sentences between the four personas. John is the junior software engineer persona, Brad is the principal software engineer persona, Harold is the English teacher persona, and Ethan is the astrophysicist persona. To generate this, we used a linguistic visual analytics framework from El-Assady et al. \cite{El-Assady2019-rg} which provides a pipeline for modelling entities within a corpus, extracting features and applying latent dirichlet allocation (LDA) for topic modelling \cite{Blei2003-zm} and then visualising the number of entities which are used in close proximity within sentences between speakers. At right is a map of the entities that were identified across all of the task prompts which illustrates the number of entities identified and their separation across different concepts.}
    \label{fig:personas-vis}
\end{figure*}

Using the framework from El-Assady et al. \cite{El-Assady2019-rg}, we can see from the visualisation in Figure 4 the relative number of entities that are common between the prompts generated by the different personas by the thickness of the lines connecting the persona names in the visualisation.  This reflects some of the results that we observed in the code, particularly the larger differences between the more technical personas - the software engineers and the astrophysicist - versus the non-technical English teacher persona.

For the generated code, across the three tasks for the different personas we observed differences between the two levels of software engineers, between them and the astrophysicist and larger differences between the software engineers and the English teacher with no programming experience, when running the experiment prompting ChatGPT. The differences in prompts for the principal software engineer emphasised deployment considerations such as scalability on large datasets and deployment in a cloud-based environment whereas the junior software engineer generally described implementation and input details. This resulted in differences in the generated code, for example, for the calculator task, providing a couple of Python classes with some tests to the junior software engineer versus the code to implement a Flask app to the principal software engineer. The astrophysicist would specify data or programming language considerations but with less emphasis on the overall system or priorities than the software engineers. The tone of the prompts differed between the software engineers and astrophysicist, which were more declarative whereas the English teacher phrased their prompts more speculatively.

The largest differences were observed between the code generated for prompts from the English teacher persona as compared with the two software engineering personas. The style and language of these prompts were the most different, with a couple of key characteristics. The English teacher persona sometimes phrased their prompts as questions such as ``can you help me with that?''; used indirect language such as ``a program that mimics a basic calculator''; and described the functionality of the artefact they were trying to build from perspective of user interactions rather than in terms of inputs and outputs, as was more common with the software engineer personas. This resulted in large differences in the generated code, with the most dramatic being no code generated when it asked for ``the computer to write up a set of financial reports'' for the accounting report generation task, which provided a set of instructions for how the user could send the reports in and it would then process them without providing a code implementation. In small ways as well, the code generated was also different between the personas, with the bicycle database example showing a password field in the user data table for both software engineer personas but not for the English teacher or astrophysicist personas.

Similar differences were observed when re-running the experiment with the same personas and prompts for Claude. An overall difference between the two models was in the amount of code produced (roughly three times more) and the granularity of the detail provided overall in Claude versus ChatGPT. For the English teacher persona versus the two software engineering personas, the code produced was more basic, such as a set of while and if/else statements for the calculator versus classes or app implementations for the software engineering personas. The code produced for the principal engineer was closer to something that could be deployed than the junior software engineer. For example, the accounting report generator for the principal software engineer provided a set of classes that would parse the data, calculate KPIs and generate the reports where as the code produced for the junior software engineer was instructions for cloning a dummy GitHub repo with an example output report and set of function calls.

The results are illustrative of our hypothesis that user background and knowledge could affect how a prompt is described and this can have downstream impacts on the code generated. This could highlight useful directions for internal and external supports to build for code LLMs, such as increased user guidance (for example, follow-up clarifying questions from the LLM or suggestions of similar code) or averaging together of variations of prompts to improve robustness (see for example\cite{Guo2024-hd}). Here we focused more specifically on software engineering experience as a key aspect of background and on problem decomposition as a key part of the description that would vary. There are, of course, other aspects of background and prompt description that should be explored in future work. This study presents first steps in investigating this challenge and future work should also explore this experiment using human participants.

\end{document}